\documentclass[global,referee]{svjour}
\usepackage{graphics}
\journalname{\apss}

\usepackage{aas_macros}

\begin{document}
\title{Active Galaxies in the UV}
\author{Wolfram Kollatschny\inst{1} \and Wang Ting-Gui\inst{2}
}                     
%
%
\institute{Institut f\"{u}r Astrophysik, Universit\"{a}t G\"{o}ttingen,
              Friedrich-Hund-Platz 1, D-37077 G\"{o}ttingen, Germany
 \and Center for Astrophysics, University of Science and Technology of China,
              Hefei, 230026, China }
%
\date{Received: 4 October 2005 / accepted 10  October 2005}
%
\maketitle
\begin{abstract}
In this article we present different aspects of AGN studies demonstrating
the importance of the UV spectral range. Most important diagnostic lines 
 for studying the general physical conditions as well as the
metalicities in the central broad line region in AGN are emitted in the UV.
The UV/FUV continuum in AGN excites not only 
the emission lines in the immediate surrounding but it is responsible for 
the ionization of the intergalactic medium in the early stages of the universe.
Variability studies of the emission line profiles of AGN in the UV give us
information on the structure and kinematics of the immediate surrounding
of the central supermassive black hole as well as on its mass itself.
\end{abstract}
\section{Introduction}
\label{intro}
Active Galactic Nuclei (AGN) are the most luminous objects in the universe.
Their luminosities, their spectral energy distribution from the radio
to the $\gamma$-ray range, as well as their emission line ratios
cannot be generated by normal stars. Galaxies containing an active 
nucleus are called active galaxies. We divide the AGN in different
subclasses such as Quasars, Seyfert galaxies and Liners.

Many aspects of the generation of the energy in AGN are still unknown.
Accretion of gas onto a central supermassive black hole (SMBH)
is generally accepted to be the dominant physical process
generating the enormous energies we are observing (Rees \cite{Rees84}).  
The accretion flow is the source of the non-thermal continuum emission
in the UV, X-ray and optical.
The spectral energy distribution (SED) of the non-thermal continuum emission
in typical AGN has its maximum in the UV. 

The central continuum source ionizes the circumnuclear gas in the so called
broad line region (BLR) and narrow line region (NLR).
The majority of the most important emission lines are emitted
in the UV spectral range.
The overall continuum distribution as well as the UV spectral lines
(narrow emission lines, broad emission lines, absorption lines) are tracers
of the physical conditions of those regions where these emission lines
originate. The emission line region of the narrow lines is spatially 
resolved in some nearby objects. They originate at distances of pc to kpc
from the central ionizing source. However, the broad emission lines originate
at distances of light days to light months only from the central ionizing
source. This BLR is unresolved by orders of magnitudes even for the nearest
AGN.   

Various excellent reviews about AGN have been published over the past
years.
Different aspects of AGN spectra were highlighted in those papers as e.g.
\cite{Netz90}, \cite{Urry95}, \cite{Kora99},\\
\cite{Hama99}, \cite{Vero00},
\cite{Ho04},\\
\cite{Heck04}, \cite{Pete04}.

This article is devoted to the UV spectral range of AGN.
The UV spectral range is important for our understanding of active galaxies
because;\\
- the maximum flux of AGN is emitted in the UV.\\
- the rest frame EUV continuum in highly redshifted AGN is important for 
our understanding of the early universe.\\
- the UV spectra of the class of low luminous AGN can only be observed in the
local universe because of their faintness.\\
-the most important diagnostic emission and absorption lines are emitted
 in the UV: they give 
information on the physical conditions in the emission line region next 
to the central ionizing source.\\
-for the study of the cosmological and chemical evolution of AGN the UV
spectra of 'nearby' objects (Z=0-2) have to be known.\\
-important far UV diagnostic lines can only be observed in the UV -- even 
for high redshift objects.\\
-variations of the emission lines give us information on the structure and
kinematics of the innermost AGN regions. The most important
lines next to the central black hole are emitted in the UV/FUV.

\section{The AGN Family}
\label{sec:1}
\subsection{Seyfert Galaxies and Quasars}
\label{sec:2}
Quasars are the most luminous subclass of the AGN family having nuclear
magnitudes of $M_{B} < -21.5$. Seyfert galaxies are by definition those
AGN with $M_{B} > -21.5$. Besides a strong non-thermal continuum their
spectra are dominated by broad permitted emission lines in the UV and
optical. Typical observed line widths (full width at half maximum (FWHM)) are 
$3000 - 6000 km s^{-1}$ with maxima of up to $30,000 km s^{-1}$. The line
widths are interpreted as Doppler motion of the BLR clouds where these
lines are emitted.
The non-thermal ionizing source in AGN is surrounded by the central
 BLR clouds at distances of less than 1 pc ($10^{15}$ to about $10^{17}$ cm).
Typical electron densities in these emission line regions are
 $n_{e}=10^{9}-10^{11} cm^{-3}$ for temperatures of about T$\sim$20.000 K.
Most of the important
 diagnostic lines of this BLR are emitted in the UV spectral range -- except
for the optical Balmer and a few Helium
 lines.

In the spectra of
Seyfert 2 galaxies only narrow (permitted and forbidden) emission lines
with typical line widths (FWHM) of $ 300 - 500 km s^{-1}$  are present  
in contrast to those of Seyfert 1 galaxies and quasars.
These  narrow emission lines originate
at distances of about 100 to 1000 pc from the center.
Electron densities in the range from $10^{2}$ to $10^{4}cm^{-3}$
are derived
 for typical electron temperatures of 10.000 - 25.000 K. Even if many
of the narrow emission lines
 are emitted in the optical wavelength range too -- the most important ones
are emitted in the UV. 
\subsection{Low Luminosity AGN}
\label{sec:5}

Low Luminosity AGN (LLAGN) refers to those objects with H$\alpha$ 
luminosities less than 10$^{39}$~erg~s$^{-1}$. They are the most abundant 
type of AGN and reside in $40\%$ of bright galaxies in the local universe
(Heckman 1980; Ho et al. 1996). There are evidence that LLAGN may consist 
of two different subclasses. The first subclass is accretion onto small 
black holes, i.e., a scaled version of Seyfert galaxies (Filippenko \& Ho 
2003; Barth et al. 2004; Greene \& Ho 2005). In the second subclass, it is 
the very low accretion rate that leads to low nuclear luminosity but 
otherwise with black holes of similar masses to those in quasars and 
Seyfert galaxies (e.g., Di Matteo et al. 2003). Both classes of objects 
have attracted much attentions in the past decade because of their role 
in the history of black hole growth in the universe and the accretion 
physics. The black hole-host galaxy connection in the low mass end of 
black hole, which is likely in their infants, is crucial to the origin 
of such relations in the massive quiescent and active galaxies, which 
were found in the last five years (e.g., Magorrian et al. 1998; 
Gebhardt et al. 2000; Ferrarese et al. 2001), and clues 
to the formation of seeded BH in the early universe. The state of very 
lower accretion rate is the end point of the AGN evolution and provides   
the test-bed for accretion process at very low rate, which is in a very 
different form from those seen in Seyfert galaxies and quasars. Very low 
radiative efficiency and the lack of big blue bump is the major prediction 
of theoretical models for the latter type (e.g., Narayan et al. 1998). 
Thus the ultraviolet observation is critical to discriminate the two 
possibilities.
 
Owing to the weakness of the active nuclei, stellar light usually dominates 
the continuum emission in the optical band even at the resolution of Hubble 
Space Telescope. As stellar spectrum drops rapidly towards ultraviolet in 
most of LLAGN, UV observation is one of the most important spectral
regimes for exploring the continuum properties of those objects. Reverberation 
mapping of broad line region described in the next section can only be 
carried out in ultraviolet for this type of AGN since one has to measure 
precisely small variations in the continuum flux.  
In addition, these AGN are so faint, only nearby objects can be studied in 
detail. However, they are much less studied in the UV than other
type of AGN due to their intrinsic faintness (Maoz et al. 1999). 

The majority of these sources show characteristics of Low Ionization Nuclear
Emission Line Region (LINER), which can be produced either through
photo-ionization of the AGN/young stellar clusters or shock process
(Heckman 1980). Some key issues that might be solved with future UV
observations include: (1) How much fraction of LINERs are powered by nuclear
activity, how much by star forming process and what is the role of shocks? 
Measuring high excitation lines (such as CIII/CII) in UV is critical to
distinguish photo-ionization process by the central continuum from opaque
shock ionization models (Dopita \& Sutherland 1996).
The  measurement of UV absorption lines of young stellar component or the
featureless AGN continuum will allow to determine the contribution of the
ionizing source, directly. (2) What is the UV continuum spectrum of these
active nuclei, which is closely related to the truncate radius of the
geometrically thin and optically thick part of the disk and coupling between
electron and proton in the case of low rate accretion onto large mass BH
(Quataert et al., 1999), or to the global energy output in the accretion onto
low mass AGN.
(3) How does the BLR structure
 of LLAGN fit into the whole picture of AGN? There is indirect evidence that
 the size of BLR in LLAGN deviates systematically from the relation
 extrapolated from the known one for Quasars and Seyfert galaxies
 (Wang \& Zhang 2003).
 But a direct measurement of the size of BLR by
 reverberation mapping is required. 
\section{Spectral Energy Distribution and rest frame EUV continuum in AGN}
\label{sec:4}
The mean broadband continuum spectral energy distribution (SED) for radio-quiet
and radio-loud AGN is shown in Fig.1. The flux scale has been normalized
at 1 $\mu$m.
\begin{figure} 
\resizebox{1.10\textwidth}{!}{%
  \includegraphics{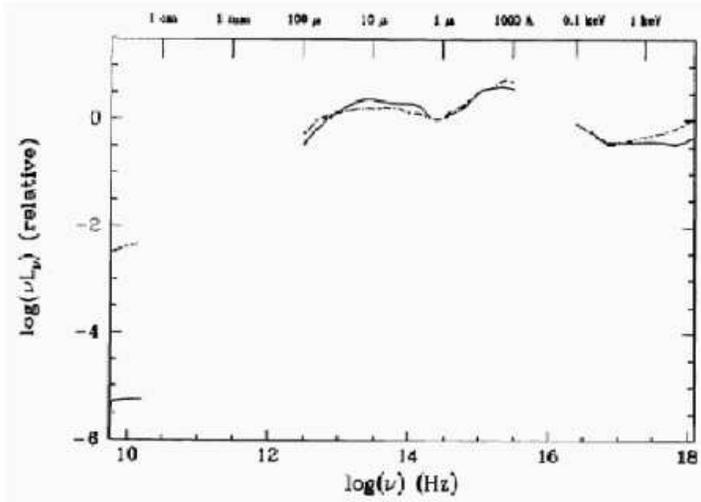}
}
%
\caption{
Schematic representation of the mean spectral energy distributions (SED)
for a sample of radio-quiet (solid lines)
 and radio-loud (dashed lines) QSOs (from \cite{Elvi94})
.}

\label{fig:1}       
\end{figure}
 The AGN continuum flux is 
relatively flat from the radio to the X-ray range. The bulk of this
flux is thought to arise from synchrotron emission.
 Besides a bump in the
infrared due to thermal dust reemission
the overall continuum flux peaks additionally in-between the optical
 and soft X-ray spectral range in the UV.
 This spectral feature is sometimes called the big blue
 bump. More than half of the bolometric luminosity of an
(un-obscured) AGN is emitted in this big blue bump.
The big blue bump is thought to arise from an accretion disk surrounding
the central black hole. Gravitational energy from the central accretion flow
is converted into the observed UV radiation
of the disk. The thermal emission in the UV corresponds to
 typical temperatures of $10^{5}$K
(e.g.\cite{Kora99}). 

 The study of the UV/EUV
 spectral range is very difficult because of the
 absorption caused by
 our own galaxy, the intrinsic absorption in distant galaxies, as
 well as the absorption in the intergalactic medium. Fig. 2a shows the
 UV composite spectrum derived from more than 2000 AGN spectra.
Before combining the spectra \cite{Telf02} corrected them for internal and
 external extinction as good as possible.
The dotted line shows accretion disk models of \cite{math87}.
The dashed line corresponds to simple power law models with a thermal 
cutoff corresponding to a temperature of $5.4~10^{5}$K.
 Fig. 2b again shows a composite
 optical--soft X-ray spectrum for radio-loud and radio-quiet quasars. One can
 see the flux is peaking not as extreme as model
 calculations of accretion disk models predict (e.g. \cite{Laor97}) . 
\begin{figure*}
\resizebox{1\textwidth}{!}{%
  \includegraphics{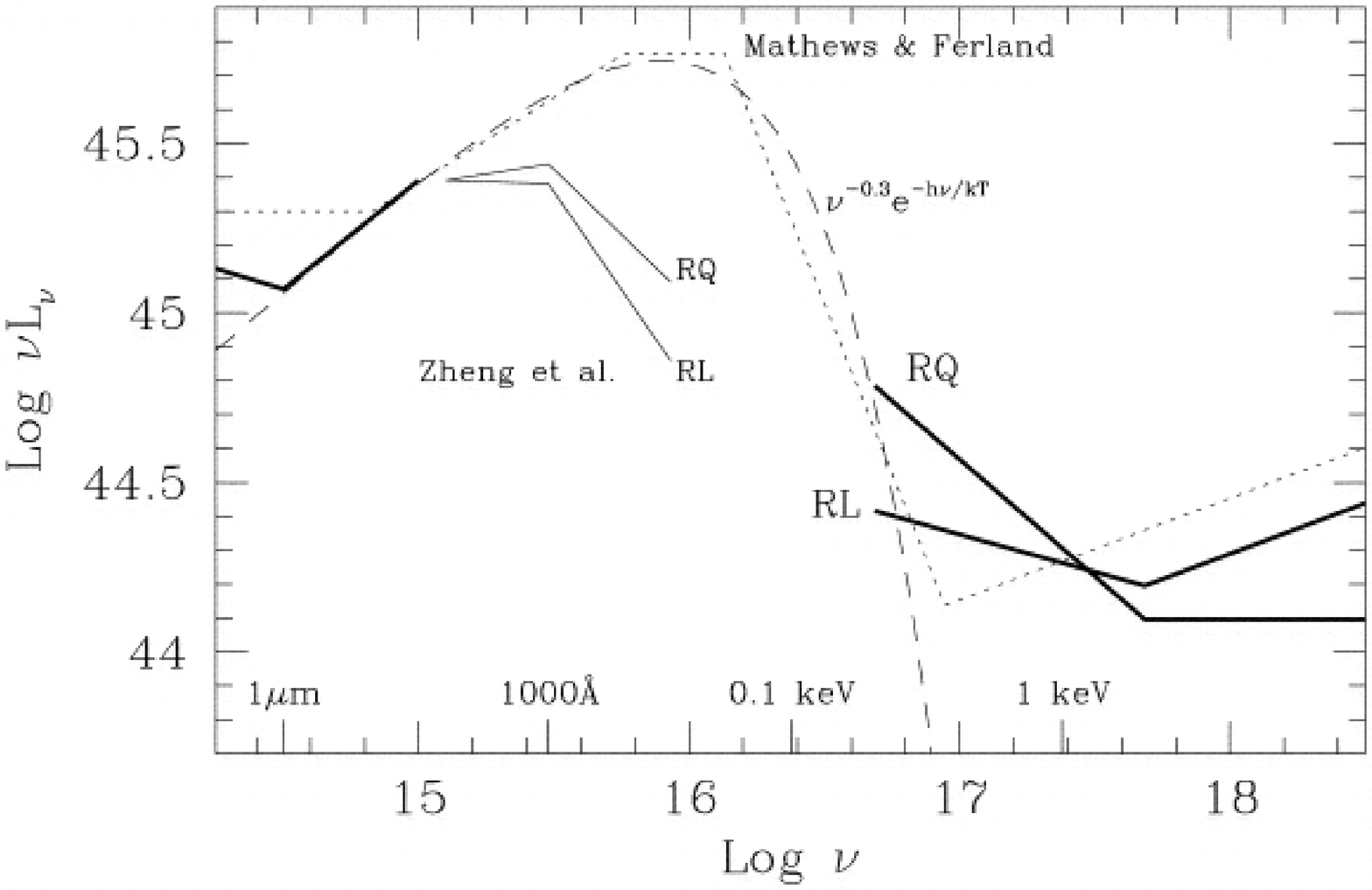}
  \includegraphics{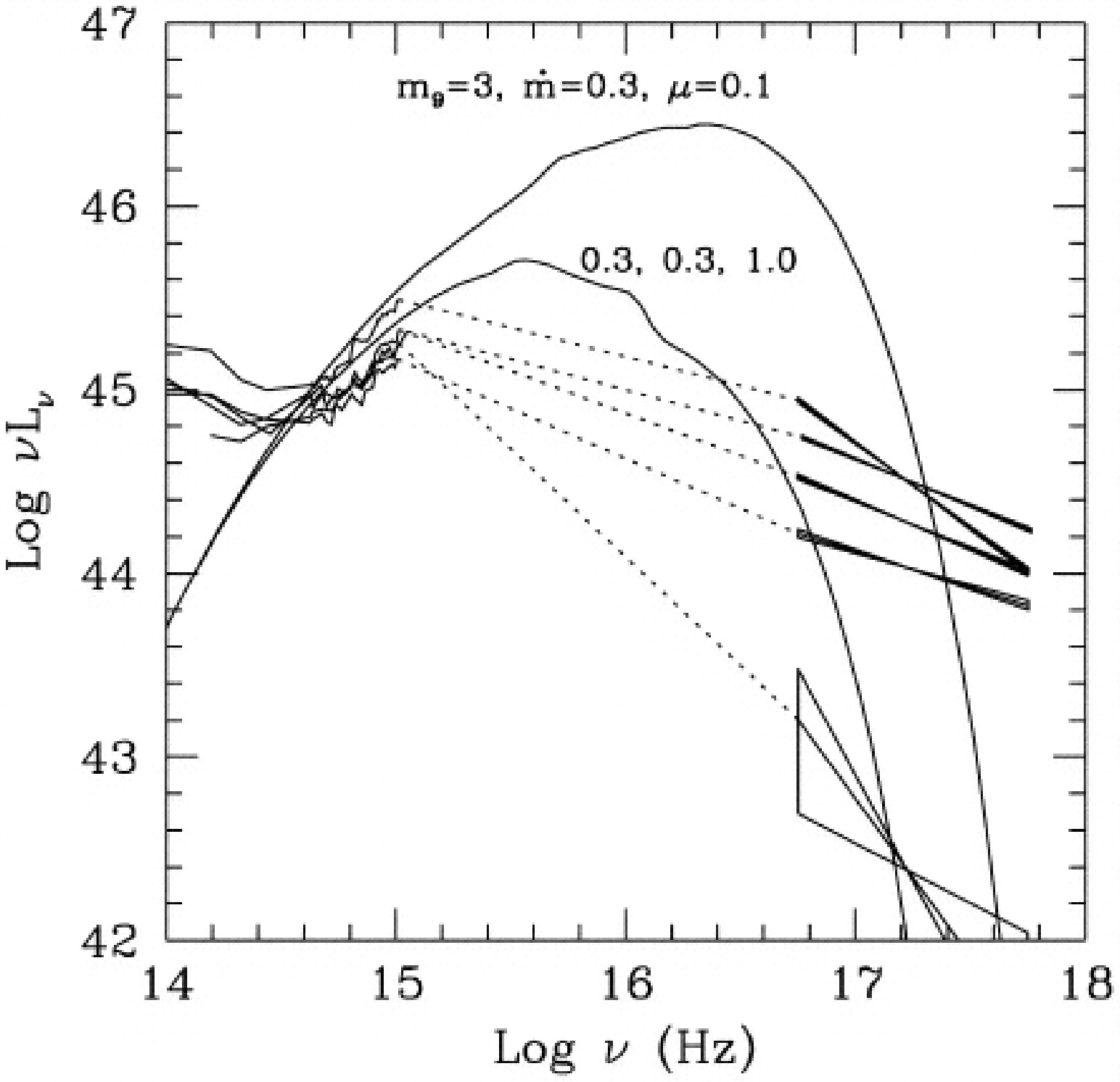}
}
\caption{Left: Composite optical--soft X-ray spectrum for the RQQs and RLQs
 in our sample (thick solid line). Three X-ray--weak quasars, and PG 1114+445,
 which is affected by a warm absorber, were excluded from the
 composite.\newline 
Right: Observed energy distribution of some quasars vs. two accretion disk
 model spectra (Laor et al., 1997).}
\label{fig:2}       
\end{figure*}
The accretion disk models cannot
 reproduce in a
simple way
 the observed spectral shape. There are indications in the
observed composite AGN spectra
 that the spectral index brakes at $\sim$ 1000 $\AA$. 
Observational difficulties are caused by dust obscuration and the contamination
of the host galaxy. Furthermore, the composite spectrum
has been derived from different classes of AGN.   
Far more observations in the UV of all classes of AGN
 are needed to understand
the details of accretion disks surrounding the central black hole in AGN.

 The knowledge of the UV/FUV spectral shape of quasars is of outmost
 importance for our understanding of the evolution of the
 early universe. The UV continuum of quasars
 ionizes the intergalactic medium at the end of the dark ages. At z$\geq$6 the
 neutral hydrogen has been re-ionized by the ionizing radiation of quasars at
 very early stages of the universe.
\begin{figure}
\resizebox{0.90\textwidth}{!}{%
  \includegraphics{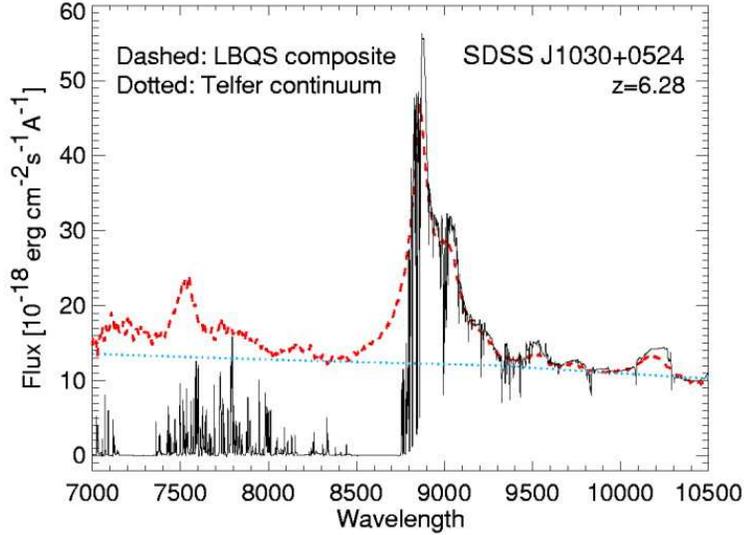}
}
%
\caption{De-noised, full-resolution spectrum of SDSS J1030+0524 with matched
 templates from the LBQS and Telfer et al. (2002. The template is a very
 good match to the quasar
 redward of the $Ly\alpha$ IGM absorption (White et al., 2003).}
\label{fig:3}       
\end{figure}
The epoch of the ionization of HeI and HeII is even less clear. Fig. 3 shows
 a spectrum of the high redshift quasar SDSSJ1030+0524 (z=6.28) with the
 UV spectral template of \cite{Telf02}.
 The Gunn-Peterson absorption troughs
 show no emission over a redshift interval of 0.2 starting at z=6. 
\section{UV emission line diagnostics}
\label{sec:5}
A UV spectrum of the Seyfert 1 galaxy NGC~4151 is shown in Fig.4.
\begin{figure}
\resizebox{1.0\textwidth}{!}{%
  \includegraphics{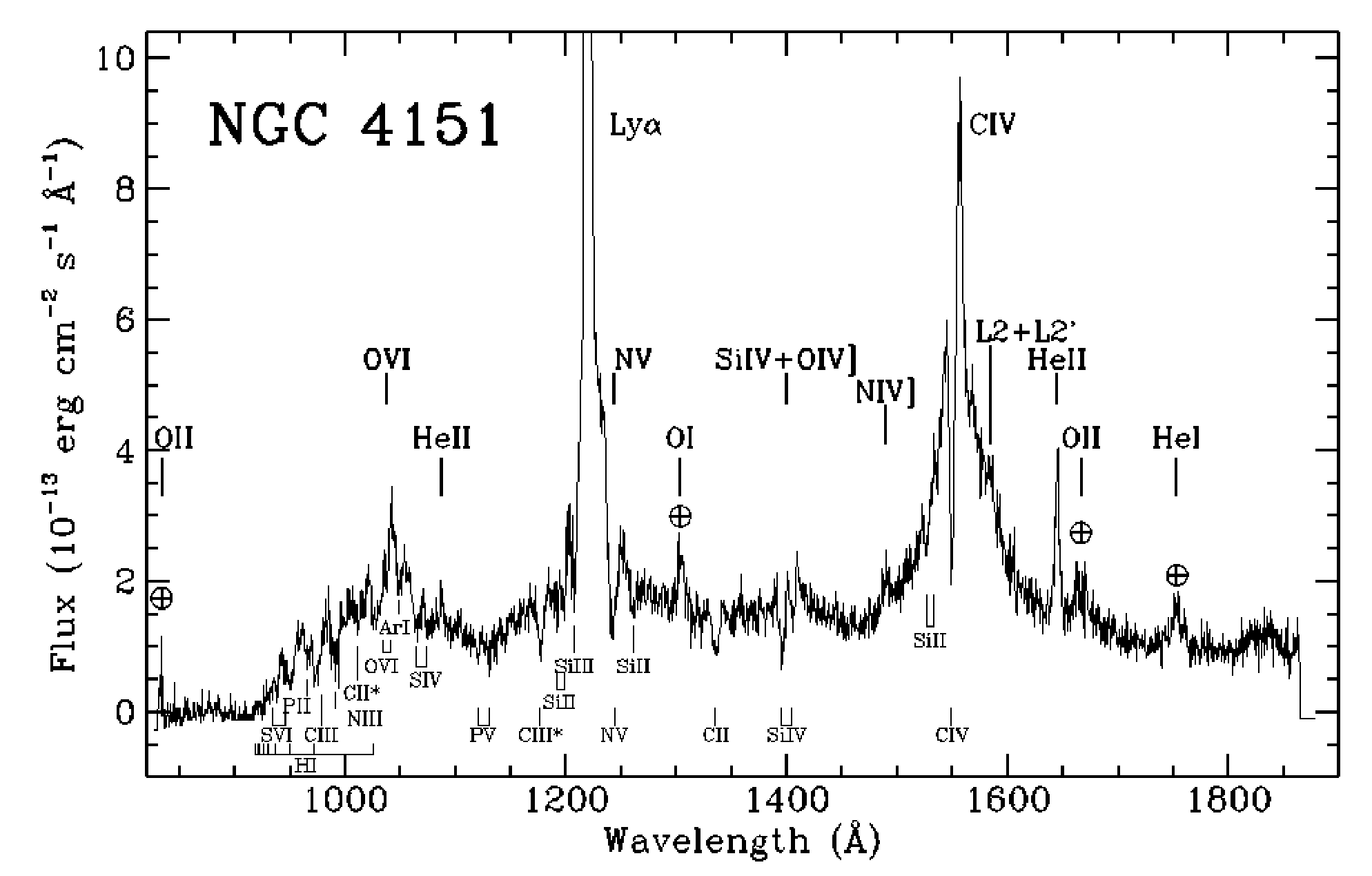}
}
\caption{The ultraviolet spectrum of the Seyfert galaxy NGC 4151 obtained by
the Hopkins Ultraviolet Telescope
(HUT).  Emission line features and absorption
features are marked.
They are due to various ionization states of different elements in the
hot gas present in the nucleus of this active galaxy
(from Kriss et al., 1992).}
\label{fig:4}       
\end{figure}
Some emission lines as well as some absorption features
are indicated in Fig.~4. The spectrum has been taken with the 
Hopkins Ultraviolet Telescope (HUT) (Kriss et al., 1992).
The most important AGN diagnostic lines between 950 and 2000\~ \AA are:
CIII 977, NIII 991, Ly$\beta$+OVI 1034, Ly$\alpha$, NV 1240, OI 1303, CII 1335,
SiIV+OIV] 1394,1402, NIV] 1486,
CIV 1549, HeII 1640, OIII] 1663, NIII] 1750, and CIII] 1909.
These emission lines show a wide range of ionization states. They originate at 
different distances from the central ionizing source in clouds 
with densities from $n_{e}=10^{8}-10^{12} cm^{-3}$.

Photoionization calculations predict line flux ratios we can compare with
the observations. Fig.~5 shows a series of calculations of emission line
ratios for different slopes of the ionizing continuum flux 
(from Hamann \& Ferland, 1999).
\begin{figure}
\resizebox{1.0\textwidth}{!}{%
  \includegraphics{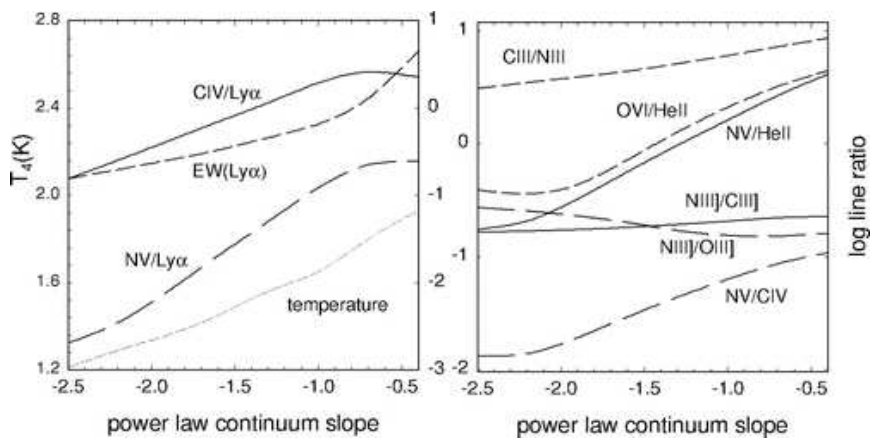}
}
\caption{
Predicted line flux ratios, gas temperatures
and dimensionless equivalent widths in Ly$\alpha$  plotted for clouds
photo-ionized by different power-law spectra.
(from Hamann \& Ferland, 1999)}
\label{fig:5} 
\end{figure}
\subsection{Metalicities}
\label{sec:6}
The determination of the
heavy element abundances in AGN is one further
aspect of quasar emission line studies.
 This is connected with the investigation of the chemical
 evolution of the universe as quasars can be observed at extreme distances
and therefore at very large look-back times. Surprisingly, the broad line
spectra of nearby AGN resemble those of the most distant quasars.
Furthermore, there are indications in the spectra of some distant
luminous quasars
that their metalicity abundances are very high even at z$\geq$5 
(e.g. \cite{Ferl96}).

In early investigations of AGN spectra the collisionally excited
inter-combination lines NIII]$\lambda$1750, NIV]$\lambda$1486,
OIII]$\lambda$16664, CIII]$\lambda$1909
 have been used to derive the
 abundance ratios of the elements nitrogen, oxygen and calcium
(e.g.\cite{Shie76},\cite{Davi77},\cite{Bald78}. But these
diagnostic lines are weak in most spectra. Furthermore,
 the densities in the BLR 
 ($n_{e}=10^{9}-10^{11} cm^{-3}$) are near the critical densities
of these lines. Therefore
these lines have different degrees of collisional
suppression.

 Permitted lines might be better candidates
for deriving the element abundances in AGN. Detailed
calculations have been carried out (e.g.\\
 \cite{Hama02})
proving the sensitivity of the UV broad emission lines
with respect to the metalicity in AGN spectra.
The most important diagnostic lines are
NIII$\lambda$991, NV$\lambda$1240, CIII$\lambda$977, CV$\lambda$1550,
CIV+OIV$\lambda$1034, HeII$\lambda$1640.

All these diagnostic lines are emitted in the UV.
It is possible to derive the metalicities
only for distant (z$\geq$2) as well as luminous quasars
when the diagnostic lines are shifted into the optical range. Very few is
 known about nearby and/or low luminous AGN. But it is necessary
 to have this information
for deriving the chemical evolution
 of the universe.

A few very interesting AGN show clear indications of abundance anomalies as
 e.g. Q0353-383 (\cite{Osme80}).
But they are rare and nothing is known about their
 evolution and their number in the present day universe. 
Very recently \cite{Bent04} checked the
 Sloan Digital Sky Survey for all nitrogen-rich quasars.
 They investigated more than 6000 quasars with
 appropriate 
redshifts that the important UV diagnostic lines were shifted into the optical
 range. Only four candidates show very strong nitrogen emission
lines comparable to those in the spectrum
of Q0353-383 (see Fig.6). This means that only about one in 1700 distant
 quasars (z$\geq$2) has extreme nitrogen over-abundances.
\begin{figure}
\resizebox{0.75\textwidth}{!}{%
  \includegraphics{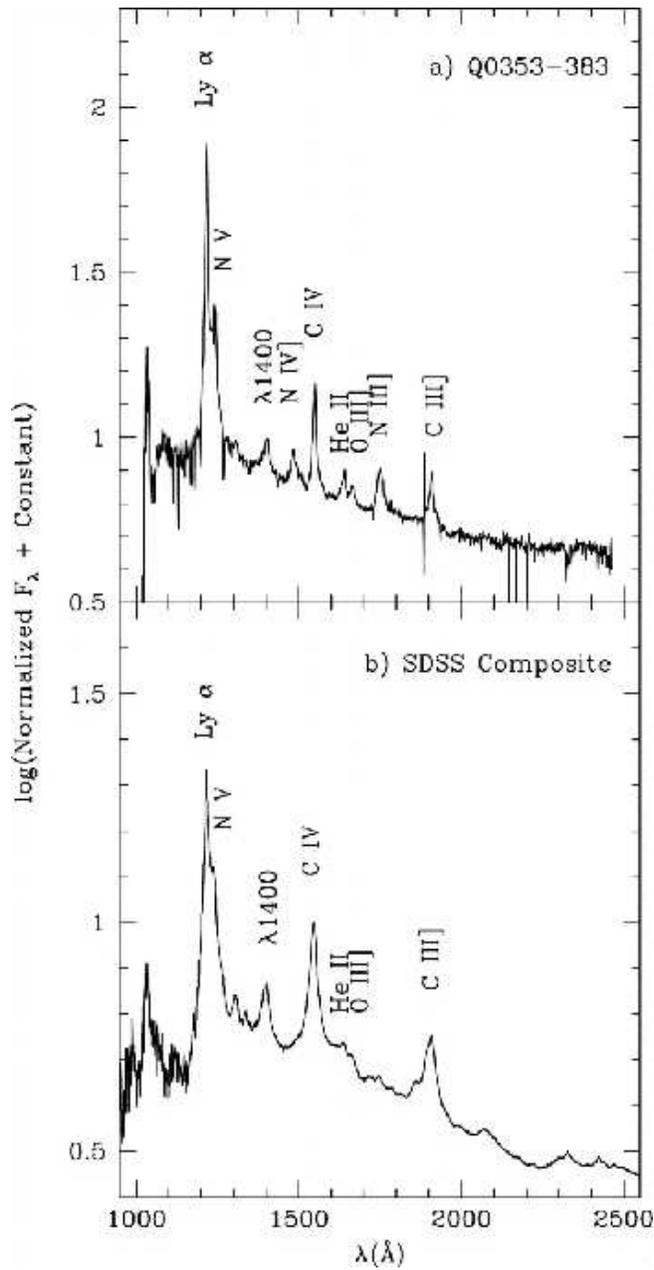}
}
%
\caption{Rest-frame spectra of (a) Q0353-383 and (b)
the SDSS composite, composed
 of 2204 quasar spectra. Both spectra are plotted in semi-log format to enhance
 fine details (Bentz et al., 2004).}
\label{fig:6}       
\end{figure}
Further spectra of nearby and distant, as well as of bright and low luminous
 AGN are needed to understand these galaxies within the overall AGN
 population. There is the basic
 question whether the nitrogen enrichment is a
 short phase in an AGN lifetime only
 or whether only a certain percentage of quasars
 reaches extremely high metalicities. We need UV spectra to detect high or
 even very high metalicities in present day AGN
to answer this question.

\subsection{Far UV diagnostic lines}
\label{sec:7}

Very few is known about line strengths of
 diagnostic emission lines in the extreme ultraviolet
 spectral range between 300 and 900 $\AA$. Composite far UV spectra
have been constructed from the spectra of highly
 redshifted QSOs taken with the Hubble Space Telescope (HST) and the Far
 Ultraviolet Spectroscopic Explorer (FUSE). They show the
 HeII$\lambda$304 and HeI$\lambda$584 lines as well as the high ionization
 NeVIII+OIV lines at 772$\AA$ and OIII at 831$\AA$
 (\cite{Telf02}, \cite{Scot04}) (see Fig.7, Telfer et al.).
\begin{figure}
\resizebox{0.90\textwidth}{!}{%
  \includegraphics{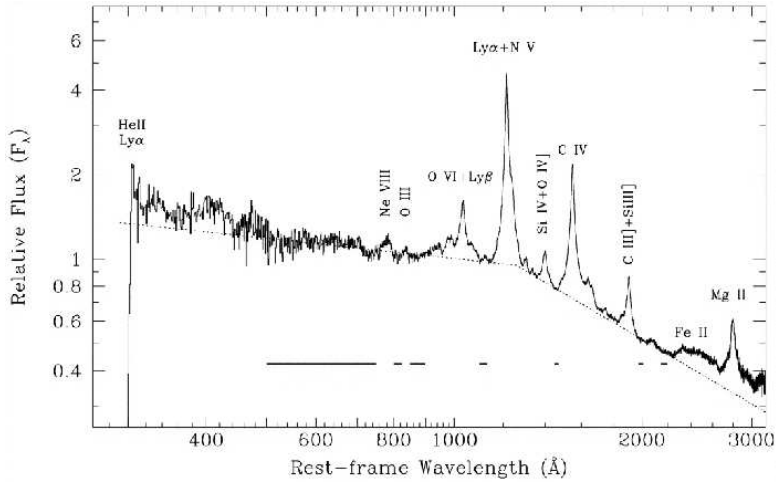}
}

\caption{Overall mean composite QSO spectrum in 1$\AA$ bins with some
 prominent emission lines marked. The dotted line shows the best-fit broken
 power-law continuum, excluding the region below 500$\AA$. The lines at the
 bottom indicate the continuum windows used in the fit (Telfer et al., 2002).}
\label{fig:7}       
\end{figure}
Considerably more UV spectra of intermediate and high redshift AGN
are needed
to compile far UV spectra with better S/N ratio and to investigate
the spectral details of different classes of AGN.

The UV and EUV diagnostic lines are of outmost importance to understand the
 AGN phenomenon. Their properties reflect
the highest energetic areas next to the central
 black holes in AGN.

\subsection{UV Absorption lines}
\label{sec:8}

Broad blue shifted resonant absorption lines in ultraviolet have been 
detected in 10-20\% optically selected quasars (\cite{wey91}), 
while narrow intrinsic absorption lines are much more common ($\sim$ 
40\% of Seyfert galaxies and $\sim$20-30\% in quasars; \cite{ham04} 
and references therein). The predominance of blue-shift among absorption 
lines suggests that partially ionized gas outflows from the active nucleus. 
Recent X-ray observations with moderate spectral resolution have found 
similar blue-shifted absorption lines in the X-ray bands (e.g., \cite{col01}
). There are suggestion that the mass loss rate and kinetic energy 
associated with the outflow may be large and can have significant impact 
on the structure of disk itself if it is disk-wind and on the ISM of the 
host galaxies. But evidence for this is still ambiguous for 
following reasons. Because strong UV absorption lines may be severely 
saturated and partially covering, the column density and ionization state 
of major UV absorbing ions are poorly determined (e.g., \cite{ara01}). 
Although the total 
absorption column density can be better determined from photo-electronic 
absorption in X-rays, very little information about velocity structure of 
X-ray absorption line/edge can be obtained from the current data. Resonant 
line absorptions in X-ray can be a very powerful diagnostics of properties 
of ions at different level of ionizations, but spectral resolution comparable 
to those in optical and UV band will not be available within next decade. 
Therefore, it is necessary to observe the weak absorption lines of the same 
elements that produce strong absorption lines in order to derive both the 
covering factor and column density as a function of velocity. Most these 
lines fall in the spectral domain of far to extreme ultraviolet.
 Figs.8,9 shows absorption lines in the UV spectral range of
3C191 taken with Keck (Hamann et al., 2001) and of Mrk509 taken with FUSE
(Kriss et al., 2003).  
\begin{figure}
\resizebox{1.00\textwidth}{!}{%
  \includegraphics{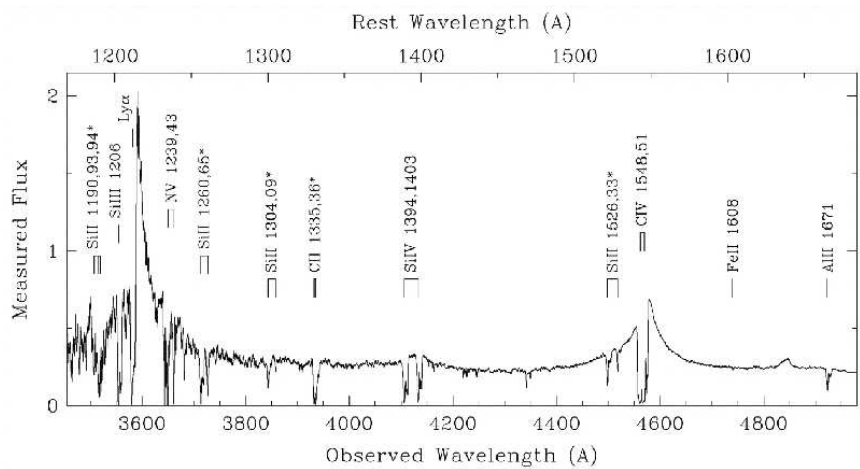}
}
%
\caption{High resolution Keck spectrum of 3C191 showing
the strong associated absorption lines in the UV
(Hamann et al., 2001).}
\label{fig:8}       
\end{figure}
\begin{figure}
\resizebox{0.6\textwidth}{!}{%
  \includegraphics{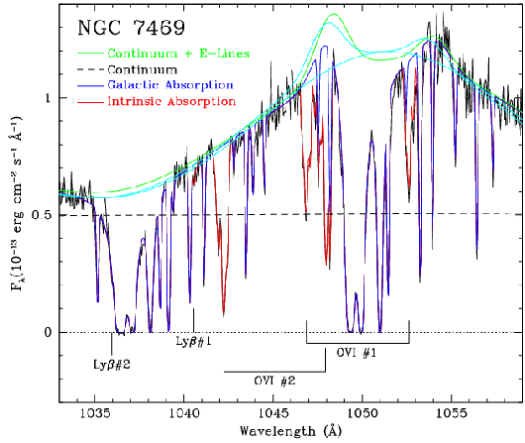}
}
%
\caption{FUSE spectrum of NGC 7469 in the Ly$\beta $/O VI region
(thin black line) (Kriss et al., 2003).}
\label{fig:9}       
\end{figure}
UV observations, simultaneously in soft X-rays with future more sensitive
 X-ray missions, may improve our understanding of the problem in several 
aspects: (1) Simultaneous observations of UV and soft X-ray absorption  
of low red-shift AGN would allow to determine the total column densities of 
material, especially those at the ionization level similar to that of UV 
absorbing material, and ionization states of the X-ray absorbing material 
(Wang et al. 2000).
At the same time we get velocity structure information from UV absorption 
lines. This will permit a detailed modeling of the physical state of outflows. 
(2) By studying absorption lines in UV bright $z\sim$2 BAL QSOs, we will 
obtain the kinematical properties of absorption lines of highly ionized 
species at far UV. Comparison of those with low ionization species will 
allow to study the changes in the kinematics with ionization state, thus 
to bridge the gap between that with X-ray absorbing material in these 
objects. Observing bright $z=2$ BAL QSOs will also allow to better determine 
the shape of the ionizing continuum, one uncertainty in the modeling of 
the ionization structure of absorbing gas. (3) Variations of intrinsic UV 
absorption lines can put strong constraints on the density of the absorbing 
material, and thus give an upper limit on the distance to the continuum 
source. If these observations are carried out simultaneously in soft X-rays 
for low-z AGN, one might distinguish the variations caused by changes in the 
flow and ionization effect (e.g.,\cite{geb03}). (4) Comparison of abundances 
derived from absorption lines with those from emission lines will give an 
independent check of those derived from emission lines.

\section{Structure and kinematics of the central region in AGN}
\label{sec:9}

The innermost line emitting region in AGN -- the broad emission line region
 (BLR) -- surrounds the central supermassive black hole at distances of
 about $10^{15}$ to $10^{17}$cm. This corresponds to radii of light days to
 light months. The motions of the line emitting clouds give us information
 on the mass of the central black hole
(e.g. \cite{Kasp00},\cite{Pete04}).
 The broad-line region is
 spatially unresolved even in the nearest AGN. But we
 can derive the structure and kinematics with indirect methods
 by studying their line and continuum variability
(e.g. \cite{Koll03},\cite{Horn04}.
\subsection{Reverberation mapping}
\label{sec:10}

In a first step one has to correlate observed light-curves of integrated
 broad emission line intensities with the ionizing continuum light curve.
It is of great advantage to observe the ionizing flux in the UV 
since the optical continuum flux is far more contaminated by the stellar
continuum flux of the host galaxy.
\begin{figure}
\resizebox{0.75\textwidth}{!}{%
  \includegraphics{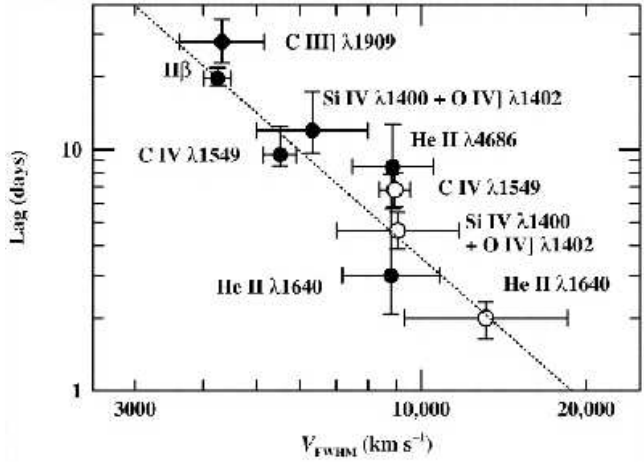}
}
%
\caption{Time lags (cross-correlation function centroids $\tau_{cent}$) in
 days (1 lt-day = $2.6 \times 10^{15} cm$) for various lines in NGC 5548 are
 plotted as a function of the FWHM of the feature (in the rest frame of
 NGC 5548) in the rms spectrum.  The
 filled circles refer to data from 1989, and the open circles refer to data
 from 1993. The dotted line indicates a fixed virial mass
 $M = 6.8 \times 10^7 M_{\odot}$ (Peterson \& Wandel, 1999).}
\label{fig:10}       
\end{figure}
Fig. 10 shows the results of an optical/UV variability campaign (including HST
observations) of the prototype Seyfert galaxy NGC5548 (\cite{Pete99}).
Plotted is the time lag of the emission lines with respect to continuum
 variations as a function of their linewidth (FWHM) in the rms profiles. The
 time lag corresponds to
 the mean distance of the line emitting region from
 the central ionizing source. One can see a clear trend: the
higher
 ionized lines originate closer to the central source. UV lines originate
about ten times closer to the center than optical emission lines.
The most successful
monitoring campaign of the integrated UV lines of an AGN
 has been carried out for NGC5548 so far (\cite{Clav91}, \cite{Kori95}).
Variability campaigns of e.g. 3C390.3 (\cite{Obri98}) and Akn~564
(\cite{Coll01}) demonstrated the power of UV reverberation studies 
but the S/N ratio and/or the fractional variability amplitudes of the
continuum variations were not strong enough for detailed line profile 
studies.

 Future monitoring campaigns of many galaxies including the UV spectral range
 of the highly ionized OVI lines
 ($\lambda\lambda$1032,1038) e.g. will uncover the innermost
 broad line region in AGN.
\begin{figure}
\resizebox{.7\textwidth}{!}{%
  \includegraphics{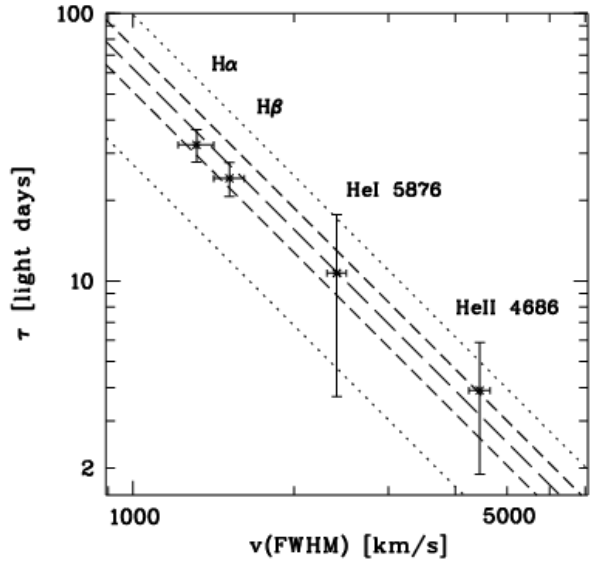}
}
\caption{The distance of the Balmer and Helium emitting line regions
from the central ionizing source in Mrk\,110
as a function of the FWHM in their rms line profiles.
The dotted and dashed lines are the results from model calculations
for central masses of 
0.8, 1.5, 1.8, 2.2, and 2.9 $\cdot10^{7} M_{\odot}$ (from bottom to top
)(Kollatschny, 2003).}

\label{fig:11}       
\end{figure}
The clear trend that higher ionized emission lines originate closer to the
 center has been seen in optical variability campaigns of e.g. Mrk 110 too
 (see Fig.~11)(\cite{Koll03}).
But the most important lines for reverberation studies are:
CV$\lambda$1550, SiIV+OIV]$\lambda$1400, HeII$\lambda$1640,
NV$\lambda$1240, CIV+OIV$\lambda$1034 (see Fig.10).
These lines give us information about the immediate surrounding
of the central black hole one order of magnitude closer than we can do
it with optical lines.

The line profile variations of UV lines should be studied
in a second step.
They gives us information on the kinematics in the broad line region. 
Detailed profile variations have been studied in the optical lines of
Mrk 110 (\cite{Koll02},\cite{Koll03}) only so far.
Different delays of emission line segments (the velocity-delay maps) measure
the geometry and flow of the line emitting gas
when we compare observed
two-dimensional velocity-delay maps with model calculations
(e.g.\cite{Wels91}).
\begin{figure*}

\resizebox{1\textwidth}{!}{%
  \includegraphics{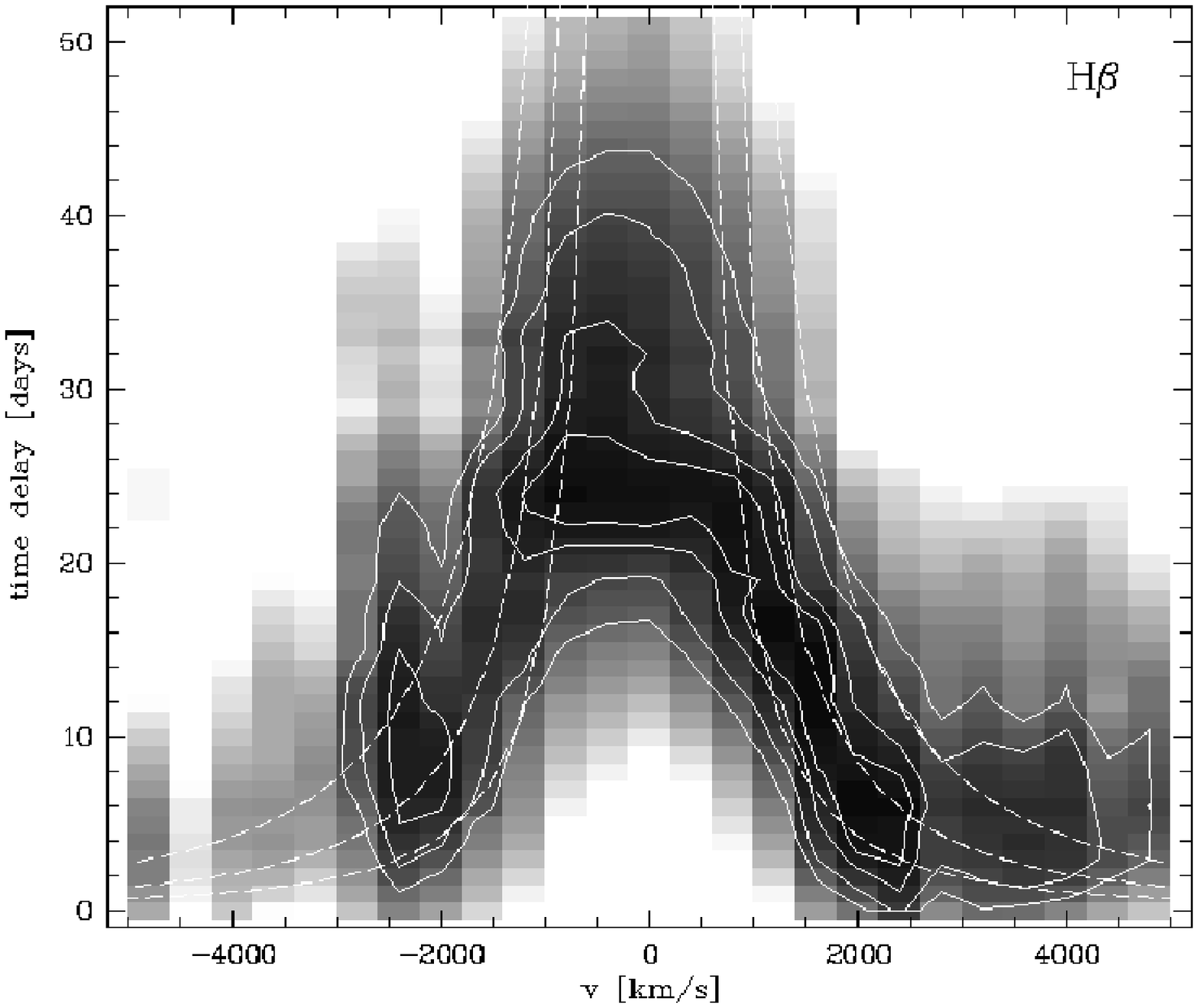}
  \includegraphics{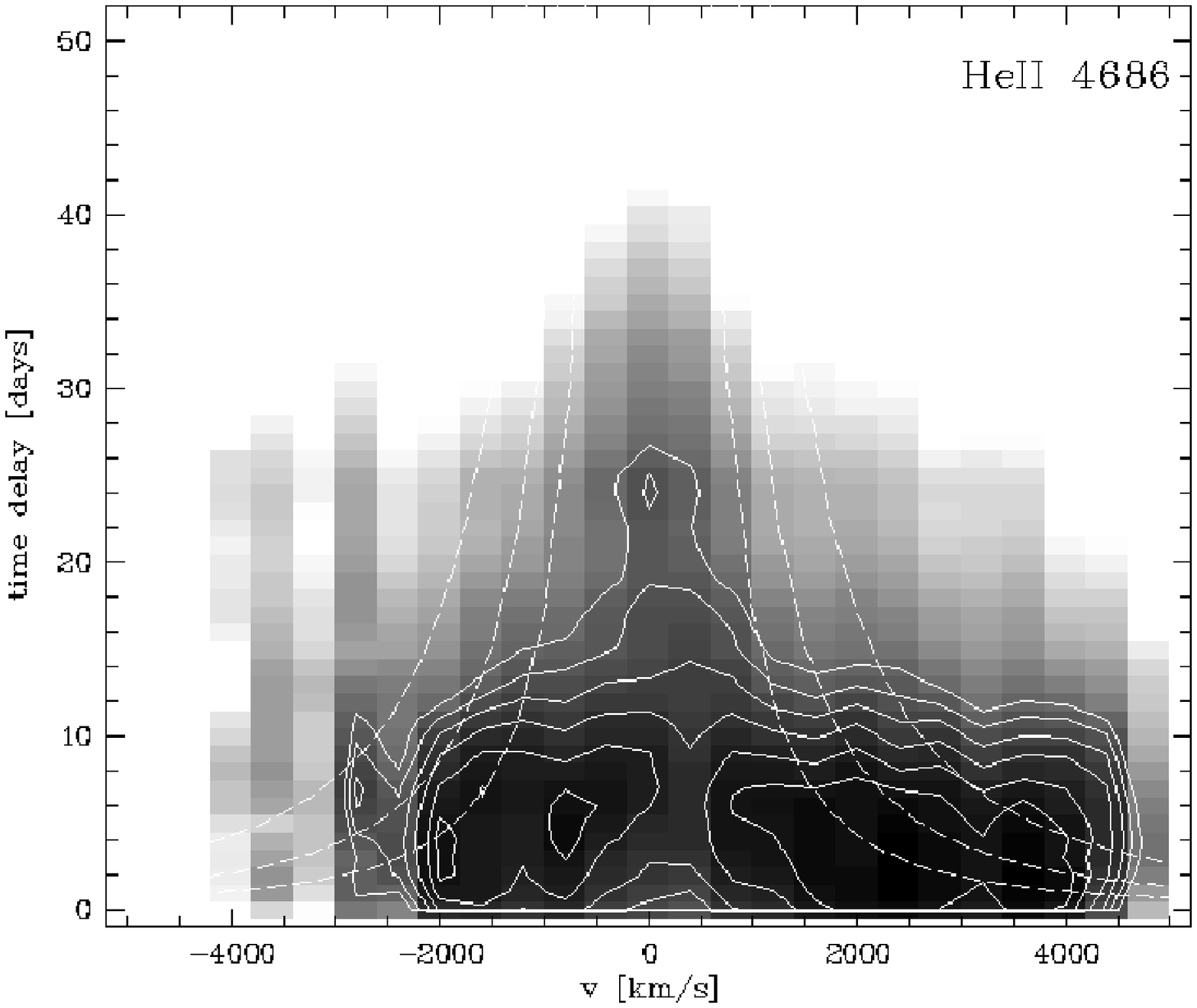}
}

\caption{The 2-D CCFs($\tau$,$v$) show the correlation of the Balmer and 
Helium line segment light curves with the continuum light curve
as a function of velocity and time delay (grey scale) in Mrk110.
Contours of the correlation coefficient are over-plotted at levels between
.800 and .925 (solid lines).
The dashed curves show computed escape velocities for
central masses of 0.5, 1., 2. $\times\ 10^7 M_{\odot}$ (from bottom to top)
(Kollatschny \& Bischoff, 2002; Kollatschny, 2003)}
\label{fig:12}       
\end{figure*}
Fig.~12 shows the correlation of H$\beta$ and HeII$\lambda$4686
 line profile segments with continuum variations.
The data are from the variability campaign
 of Mrk 110 taken with the 10m Hobby Eberly Telescope at McDonald Observatory.
 Only Keplerian disk BLR models can reproduce the
observed fast and symmetric response
 of the outer line wings. The  H$\beta$ line center
 originates at distances of
 25 light-days while the HeII line center originates at distances
of 4 light-days only. 

\subsection{Central Black Hole mass in AGN}
\label{sec:11}

It is possible to calculate the central black hole mass in AGN.
One has to
know the distances of the line emitting clouds as well as
the velocity dispersion of these clouds (e.g. \cite{Pete04}).
We derived a central black hole mass of $1.4*10^{8}M_{\odot}$ in Mrk110.
In that case we used
additional information about the projected angle of the accretion disk where
the broad emission lines originate \cite{Koll03}.
 A Schwarzschild radius $r_s$ of
 $4*10^{13}cm$ corresponds to this black hole mass. 
\begin{figure}
\resizebox{0.75\textwidth}{!}{%
  \includegraphics{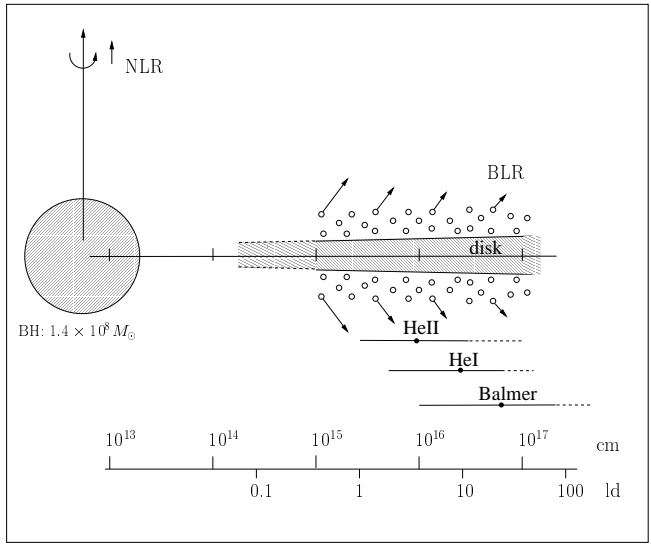}
}
%
\caption{Schematic model of the innermost region in the Seyfert galaxy Mrk110
derived from 2D-reverberation mapping. (Kollatschny, 2003).}
\label{fig:13}       
\end{figure}
Fig.~13 shows the inner broad line region structure of Mrk 110
derived from 2D-reverberation mapping.
 The HeII line originates at a distance of 230 Schwarzschild radii only
 from the
 central black hole. The monitoring of highly ionized UV lines in AGN 
enables us to study the physics of the immediate environment of black holes
 even more closer to the center. This helps us to derive
the central black hole mass more precisely.
Finally, we will achieve a clear progress
 in our knowledge of black hole physics
by monitoring different types of AGN in the UV.

\clearpage



\begin{thebibliography}{}

\bibitem[Arav et al. 2001]{ara01} Arav, N., et al., 2001, ApJ, 561, 118

\bibitem[Baldwin \& Netzer(1978)]{Bald78} Baldwin, J.~A.~\& 
Netzer, H.\ 1978, \apj, 226, 1 

\bibitem[Barth et al.(2004)]{Bart04} Barth, A.J., Ho, L.C., Rutledge R.S.,
 Sargent W.L.C., 2004, ApJ, 607, 90

\bibitem[Bentz, Hall, \& Osmer(2004)]{Bent04} Bentz, M.~C., 
Hall, P.~B., \& Osmer, P.~S.\ 2004, \aj, 128, 561 

\bibitem[Clavel et al.(1991)]{Clav91} Clavel, J. et al.,
\ 1991, \apj, 366, 64 

\bibitem[Collinge et al. (2001)]{col01} Collinge, M.J., et al., 2001, ApJ, 557, 2

\bibitem[Collier et al.(2001)]{Coll01} Collier ,S.,Crenshaw , D.~M.,
 Peterson ,B.~M., W. N. Brandt W.N. et al., 2001, ApJ, 561, 146

\bibitem[Davidson (1977)]{Davi77} Davidson, K. 
1977, \apj, 218, 20 

\bibitem[Di Matteo et al.(2003)]{Dima03} Di Matteo,T., Allen, S.W. Fabian A.C.,
Wilson A.,Young A.,J. \ 2003, ApJ, 582, 133

\bibitem[Dopita \& Sutherland(1996)]{Dopi96} Dopita M.A., Sutherland R.S.,
 1996, ApJS,102,16

\bibitem[Elvis et al.(1994)]{Elvi94} Elvis, M., 
Wilkes, B.J., McDowell, J.C., Green, R.F. et al. \ 1994, ApJS, 95, 1 

\bibitem[Ferland et al.(1996)]{Ferl96} Ferland, G.~J., 
Baldwin, J.~A., Korista, K.~T., Hamann, F., Carswell, R.~F., Phillips, M., 
Wilkes, B., \& Williams, R.~E.\ 1996, \apj, 461, 683 

\bibitem[Ferrarese et al., 2001]{fer01} Ferrarese L., Pogge R.~W., Peterson B.~M., Merritt D., Wandel A., Joseph C.~L., 2001, ApJ, 555, L79

\bibitem[Filippenko \& Ho(2003)]{Fili03} Filippenko A.V., Ho, L.C., 2003,ApJ,
 588, L13

\bibitem[Gebhardt et al., 2000]{Geb2000} Gebhardt K., et al., 2000, ApJ, 539, 

\bibitem[Gebel et al. 2002]{geb03} Gebel J.R., et al. 2003, ApJ, 595, 120

\bibitem[Hamann \& Ferland (1999)]{Hama99} Hamann, F. \& Ferland, G.~J.
\ 1999, ARAA, 37, 487 

\bibitem[Hamann et al. (2001)]{Hama01} Hamann, F.,
Barlow, T. A.; Chaffee, F. C.; Foltz, C. B.; Weymann, R. J.\ 2001, \apj,
 550, 142

\bibitem[Hamann et al. (2002)]{Hama02} Hamann, F., Korista, 
K.~T., Ferland, G.~J., Warner, C., \& Baldwin, J.\ 2002, \apj, 564, 592 

\bibitem[Hamann \& Sabra (2004)]{ham04} Hamann, F., \& Sabra, B., 2004, in 
ASP conf Series 311, ed. G.T., Richards \& P.B. Hall, (San-Francisco ASP), 
203
\bibitem[Heckman (1980)]{Heck80}Heckman, T.M., 1980, A\&A, 87, 142

\bibitem[Heckman (2004)]{Heck04}Heckman, T.M., 2004, in Coevolution of Black
Holes and Galaxies, from the Carnegie Observatories Centennial Symposia.
Published by Cambridge University Press, ed. L. C. Ho, p. 359.

\bibitem[Ho et al. (1997)]{Ho97} Ho, L.C., Filippenko, A.V., Sargent, W.L.W.,
1997, ApJ, 487, 568

\bibitem[Ho (2004)]{Ho04}Ho, L., 2004, in Coevolution of Black
Holes and Galaxies, from the Carnegie Observatories Centennial Symposia.
Published by Cambridge University Press, ed. L. C. Ho, p. 292.

\bibitem[Horne et al. (2004)]{Horn04} Horne, K., Peterson, B.M.,
Collier, S.M., Netzer, H., 2004, PASP, 116, 465

\bibitem[Kaspi et al. (2000)]{Kasp0}
Kaspi, S., Smith, P.S., Netzer, H., Maoz, D., Jannuzi, B.T., Giveon, U.
\ 2000, ApJ, 533, 631

\bibitem[Kollatschny \& Bischoff (2002)]{Koll02} Kollatschny, 
W.~\& Bischoff, K.\ 2002, \aap, 386, L19 

\bibitem[Kollatschny (2003)]{Koll03} Kollatschny, W.\ 2003, 
\aap, 407, 461 

\bibitem[Koratkar \& Blaes (1999)]{Kora99} Koratkar, A. \& Blaes, O.\ 1999,
 PASP, 111, 1

\bibitem[Korista et al. (1995)]{Kori95} Korista, K. T., et al. 1995,
 ApJS, 97, 285

\bibitem[Kriss et al.(1992)]{Kris92} Kriss, G. A.; Davidsen, A. F.;
Blair, William P. et al.\ 1992, \aap, 392, 485

\bibitem[Kriss et al.(2003)]{Kris03} Kriss, G. A.; Blustin, A.;
Branduardi-Raymont, G.; Green, R. F.; Hutchings, J.; Kaiser, M. E.\ 2003,
\aap, 403, 473

\bibitem[Laor et al.(1997)]{Laor97} Laor, A., Fiore, F., 
Elvis, M., Wilkes, B.~J., \& McDowell, J.~C.\ 1997, \apj, 477, 93 

\bibitem[Magorrian et al., 1998]{mag98} Magorrian J., et al., 1998, AJ, 115, 2285 L13

\bibitem[Mathews \& Ferland(1987)]{math87} Mathews, W.~G. \& Ferland , 
G.~J.\ 1987, \apj, 323, 456 

\bibitem[Maoz et al.(1999)]{Maoz99} Maoz D. et al., 1999, AJ, 116, 55

\bibitem[Narayan(1998)]{Nara98} Narayan R., Mahadevan R., Grindlay J.E., Popham
R.G., Gammie C., 1998, \apj, 492, 554

\bibitem[Netzer (1990)]{Netz90} Netzer,H. 1990, in Active galactic nuclei,   
Saas-Fee advanced course 20., T.J.-L. Courvoisier et al. eds, p.57

\bibitem[O'Brien et al.(1998)]{Obri98} O'Brien, P.~T. et al.\
 1998, ApJ, 509, 163O

\bibitem[Osmer \& Smith(1980)]{Osme80} Osmer, P.~S.~\& Smith, 
M.~G.\ 1980, \apjs, 42, 333 

\bibitem[Peterson et al. (2004)]{Pete04} Peterson, B.~M., Ferrarese, L.,
 Gilbert, K.M., Kaspi, S. et al. \ 2004, \apj, 613, 682

\bibitem[Peterson \& Wandel(1999)]{Pete99} Peterson, B.~M.~\& 
Wandel, A.\ 1999, \apjl, 521, L95 

\bibitem[Quatert et al.(1999)]{Quatl99} Quataert E., di Matteo T., Narayan R.,
 Ho L.C., 1999, ApJ, 525, L89

\bibitem[Rees (1984)]{Rees84} Rees M., 1984, ARA\&A, 22, 471

\bibitem[Scott et al.(2004)]{Scot04} Scott, J.~E., Kriss, 
G.~A., Brotherton, M., Green, R.~F., Hutchings, J., Shull, J.~M., \& Zheng, 
W.\ 2004, \apj, 615, 135 

\bibitem[Shields (1976)]{Shie76} Shields, G.~A. 
1976, \apj, 204, 330 

\bibitem[Telfer, Zheng, Kriss, \& Davidsen(2002)]{Telf02} 
Telfer, R.~C., Zheng, W., Kriss, G.~A., \& Davidsen, A.~F.\ 2002, \apj, 
565, 773 

\bibitem[Urry \& Padovani(1995)]{Urry95} Urry, C.~M. \& Padovani, 
P.\ 1995, PASP, 107, 803

\bibitem[Veron \& Veron(2000)]{Vero00} Veron, M.~P. \& Veron, 
P.\ 2000, Astron.Astrophys Rev., 10, 81

\bibitem[Wang et al. (2000)]{wan00} Wang, T.~G., Brinkmann, 
W., Yuan, W., Wang, J.~X., \& Zhou, Y.~Y.\ 2000, \apj, 545, 77
\bibitem[Wang \& Zhang(2003)]{Wang03} Wang T.G., Zhang X.G., 2003,
 MNRAS,340,793

\bibitem[Welsh \& Horne(1991)]{Wels91} Welsh, W.~F.~\& Horne, 
K.\ 1991, \apj, 379, 586 

\bibitem[White et al.(2003)]{Whit03} White, R.~L., Helfand, 
D.~J., Becker, R.~H., Gregg, M.~D., Postman, M., Lauer, T.~R., \& Oegerle, 
W.\ 2003, \aj, 126, 706 

\end{thebibliography}
\end{document}